\documentclass{iopart}

\usepackage{iopams,setstack}
\usepackage{epsf}
\usepackage{cite}

\newcommand{\bd}{\begin{displaymath}}
\newcommand{\ed}{\end{displaymath}}
\newcommand{\be}{\begin{equation}}
\newcommand{\ee}{\end{equation}}
\newcommand{\ba}{\begin{eqnarray}}
\newcommand{\ea}{\end{eqnarray}}

\begin{document}

\paper[Quantum Zeno and anti-Zeno effects in surface diffusion]
{Quantum Zeno and anti-Zeno effects in surface
diffusion of interacting adsorbates}

\author{H C Pe\~nate-Rodr\'{\i}guez$^1$, R Mart\'{\i}nez-Casado$^2$,\newline
G Rojas-Lorenzo$^1$, A S Sanz$^3$ and S Miret-Art\'es$^3$}

\address{$^1$Instituto Superior de Tecnolog\'{\i}as y Ciencias Aplicadas,
Ave.\ Salvador Allende y Luaces, Quinta de Los Molinos,
Plaza, La Habana 10600, Cuba}

\address{$^2$Department of Chemistry, Imperial College London,
South Kensington, London SW7 2AZ, United Kingdom}

\address{$^3$Instituto de F\'{\i}sica Fundamental - CSIC,
Serrano 123, 28006 Madrid, Spain}

\eads{\mailto{asanz@iff.csic.es}}

\begin{abstract}
Surface diffusion of interacting adsorbates is here analyzed
within the context of two fundamental phenomena of quantum
dynamics, namely the quantum Zeno effect and the anti-Zeno effect.
The physical implications of these effects are introduced here in
a rather simple and general manner within the framework of
non-selective measurements and for two (surface) temperature
regimes: high and very low (including zero temperature). The
quantum intermediate scattering function describing the adsorbate
diffusion process is then evaluated for flat surfaces, since it is
fully analytical in this case. Finally, a generalization to
corrugated surfaces is also discussed. In this regard, it is found
that, considering a Markovian framework and high surface
temperatures, the anti-Zeno effect has already been observed,
though not recognized as such.
\end{abstract}






\section{Introduction}
\label{sec1}

Traditionally, one finds in the literature that activated surface
diffusion and related phenomena, such as the frustrated
translational motion (the so-called $T$-mode), are well described
by classical models whenever heavy particles and long time scales
(of the order of tens and hundreds of picoseconds) are considered.
In such cases, which are the typical ones, any trace of
quantumness is essentially swept out; one only finds clues of the
lost quantumness through ``reminders'', for example, in recoil
energies. However, below these time scales also explored through
collisions between the probe particle and the diffusive one,
quantum effects should be stronger. In this domain, the
commutation rule for position operators at different times should
play an important role in the corresponding dynamics, for example.
In principle, this time domain would be ruled by a coherence time
$t_c \sim \hbar \beta$($\beta = 1 / k_B T$ and $k_B$ is the
Boltzmann constant), determined essentially by the surface
(substrate) temperature $T$. Thus, in general, for times smaller
than $t_c$, quantum effects should be dominant over thermal
effects. Hence, a question that arises in a natural way is how and
why this transition takes place or, in other words, why the
diffusing, interacting adsorbates lose their quantum behavior.
Within the scenario depicted by this quantum-to-classical
transition, one can understand surface diffusion processes as
decaying processes, with their evolution being monitored along
time through the so-called intermediate scattering function.

Transitions like the aforementioned one are strongly connected
with measure processes in the sense that, regardless of how a
measure is defined, in general such processes can be understood as
a system coupled to an external agent acting on it continuously or
at certain times.
With each interaction between the system and the measuring
device, the former would lose part of its quantumness, thus
leading to typical classical behaviors ({\it decoherence}). This
brings us to the scenario of the quantum Zeno effect (QZE) and its
complementary, the anti-Zeno effect (AZE), which consist of precisely
inhibiting or enhancing, respectively, the system natural
(quantum) decay by carrying out a series of measures on it. These
effects appear and are analyzed commonly within the context of
unstable quantum systems. However, here we would like to introduce
them within a very different context, namely surface diffusion of
interacting adsorbates. In this regard, a novel analysis and (to
some extent) new interpretation of quantum surface diffusion in
terms of QZE and AZE (excluding tunneling mediated diffusion) is
presented. The primary motivation of this work arises from the
possibility at present to carry out surface diffusion experiments
finely resolved in time by means of neutron \cite{peter} and
$^3$He spin-echo techniques \cite{allison1,allison2}. From these
experiments, one obtains the time-evolution of a quantity (namely
the so-called polarization function) proportional to the
intermediate scattering function,
\be
 I(\Delta {\bf K},t) = \langle e^{-i \Delta {\bf K} \cdot {\bf R}(0)}
  e^{i \Delta {\bf K} \cdot {\bf R}(t)} \rangle .
 \label{zeno10}
\ee
This time-dependent (correlation) function is the space Fourier
transform of the time-dependent pair correlation function or
$G$-van Hove function, $G({\bf R},t)$, a generalization of the
pair-distribution function from the theory of liquids
\cite{vanhove,vineyard}. Given an adparticle is at some position
${\bf R}(0)$ on the surface at time $t=0$, $G({\bf R},t)$ accounts
for the averaged probability of finding the same or another adparticle
at ${\bf R}$ at a time $t$ (here, ${\bf R}$ denotes the adparticle
position operators given in the Heisenberg picture). To some
extent, one could therefore interpret this function as a {\it time
decay law}, describing the loss of correlation between a system
initial configuration, represented by $e^{i \Delta {\bf K} \cdot
{\bf R}(0)}$, and its configuration at time $t$, $e^{i \Delta {\bf
K} \cdot {\bf R}(t)}$, measured along the $\Delta {\bf
K}$-direction ($\Delta {\bf K}$ is the wave-vector transfer
parallel to the surface of the probe particles, i.e., neutrons or
$^3$He atoms).

Experimentally, it is observed that $I(\Delta {\bf K},t)$ decays
smoothly with time, quadratically at short times and like
$e^{-\eta_{\rm eff} t}$ at longer times, with $\eta_{\rm eff}$
depending on the reciprocal lattice associated with the surface,
the momentum transfer and the friction coefficient $\eta$ ruling
the diffusion time scale. In activated surface diffusion, the
quantum Langevin equation formalism has been considered to
evaluate the intermediate scattering function (\ref{zeno10}). The
Caldeira-Leggett Hamiltonian model has also been used to describe
the diffusion dynamics, representing the surface as an infinite
collection of harmonic oscillators at a given surface temperature
$T$ (reservoir) and with an Ohmic (constant) friction (one-bath
model \cite{salva1}). If interacting adsorbates are considered,
the total friction splits up into two contributions, $\eta =
\gamma + \lambda$, as shown by means of the so-called interacting
single adsorbate (ISA) approximation, which has been applied with
success at low and intermediate coverages (two-bath model
\cite{ruth1,ruth2}). Within this model, $\gamma$ is associated
with the substrate or reservoir, while $\lambda$ represents a
collisional friction due to the collisions among adsorbates, thus
being connected with the surface coverage. By invoking the
elementary kinetic theory of transport of gases, a simple relation
can be found between the collisional friction and the coverage
$\theta$ at a temperature $T$
\begin{equation}
\lambda = \frac{6 \rho \theta}{a^2} \sqrt{\frac{k_B T}{m}} ,
\label{la}
\end{equation}
where $a$ is the unit cell length of an assumed square surface
lattice and $\rho$ is the effective radius of an adparticle of
mass $m$. From a quantum-mechanical viewpoint, when the adsorbate
interacts with the surface the entanglement with the degrees of
freedom of the latter leads to the loss of coherence (or
decoherence) of the former, i.e., its ability to display quantum
behaviors in the corresponding dynamics (e.g., quantum
interference) \cite{giulini}. In the context of quantum Brownian
motion \cite{weiss}, this phenomenon where a highly delocalized
state in position and/or momentum transforms into a localized
classical state is called {\it environment induced decoherence}.
In the particular case of the adsorbate surface diffusion, the
role of the measurement process on the adsorbates would be
associated with the surface and the other surrounding adsorbates.
This process thus belongs to the class of non-selective,
continuous and indirect measurements, which are different to the
more common projective (von Neumann) measurements. As a
consequence, the adsorbate position would be describable by a
Langevin equation, where the apparatus would be included through
an interaction Hamiltonian \cite{petruccione}.
More specifically, the apparatus would be a reservoir \cite{joos,hakim}, where the effect of
indirect, non-selective measurements (e.g., collisions) would
contribute to the dephasing of the system quantum state, making
the phase of the adparticle completely random \cite{kurizki}.
Within the two-bath model, the surrounding adsorbates form a
reservoir, such that their collisions with the tagged adsorbate
(and therefore their frequency, measured through $\lambda$) can be
controlled through the coverage after Eq. (\ref{la}).

Finally, we would like to mention that in the context of
atom-surface scattering, Levi has recently introduced and
discussed decoherence \cite{levi1} and the QZE \cite{levi2}.

The organization of this work is as follows. To be self-contained,
in Section~\ref{sec0} the physics behind the QZE and AZE is
introduced in a brief manner. In Section~\ref{sec2} an overview of
the wave packet diffusion dynamics through flat surfaces is
presented in order to introduce some concepts and notions on QZE
and AZE within the context of surface diffusion. We would like to
stress here that the appearance of QZE and AZE is very much
related to the type of measurement carried out, leading to
conclusions which can be quite different.
In Section~\ref{sec3}, the intermediate scattering function for flat
and corrugated surfaces is analyzed, seeing it as a decay rate in the
Heisenberg picture, and provides a new interpretation of the
experimental and theoretical results previously reported. Finally,
in Section~\ref{sec4},
some conclusions are present together with possible extensions of
this type of analysis to the frustrated translational mode.


\section{Elementary grounds of QZE and AZE}
\label{sec0}

In the standard fashion, the QZE and AZE are usually introduced within
the context of a series of ideal, repeated measurements (or
projections) on a system during its time-evolution
\cite{misra,peres,joos,milburn,home,gontis,itano,kofman,kurizki,fischer,%
facchi,pascazio}.
Thus, let us first focus on the short-time decay
dynamics of an unstable system described by the total
Hamiltonian
\be
 H = H_0 + V ,
 \label{zeno1}
\ee
where $H_0$ accounts for the system free dynamics and $V$ describes
some interaction that ultimately leads to the system decay.
At $t = 0$, the system is supposed to
be in a pure state $|\psi_0 \rangle$, which is a normalized vector
in Hilbert's space. The {\it survival probability} or probability
to find the system in the same state at a certain subsequent time
$t$ is given by
\be
 P(t) = |\langle \psi_0| \psi(t)\rangle|^2
      = |\langle \psi_0| e^{-iHt/\hbar} |\psi_0\rangle|^2 .
 \label{zeno2}
\ee
At very short times, $P(t)$ can be expressed by Taylor expansion
of the evolution operator up to second order as
\be
 P(t) \approx 1 - \left(
  \langle \psi_0| H^2 |\psi_0\rangle -
  \langle \psi_0| H |\psi_0\rangle^2 \right)
  \frac{t^2}{\hbar^2}
  = 1 - \left( \frac{t}{\tau_Z} \right)^2 .
 \label{zeno3}
\ee
That is, at very short times the survival probability decays
quadratically with time. If initially the system is in an
eigenstate of $H_0$, then we find the characteristic Zeno time
$\tau_Z$ depends essentially on the variance of the interaction
potential, i.e.,
\be
 \tau_Z = \frac{\hbar}{\sqrt{
  \langle \psi_0| V^2 |\psi_0\rangle -
  \langle \psi_0| V |\psi_0\rangle^2 }} .
 \label{zeno4}
\ee
Now, consider a number $N$ of instantaneous ideal measurements
(projections) are performed at very small intervals of time $\tau$,
such that $t = N\tau$, in order to ascertain whether the system still
remains in its initial state or not.
Each time a measurement is
performed, the system wave function is ``collapsed'' and its
subsequent evolution starts again from the state $|\psi_0\rangle$.
Therefore, the probability to find the system in this state
after a time $t$ and after performing $N$ measurements will be
\be
 P^{(N)}(t) = \left[ P(\tau) \right]^N
  = \left[ 1 - \left( \frac{t}{N\tau_Z} \right)^2 \right]^N .
 \label{zeno5}
\ee
In the limit $N \to \infty$, this quantity becomes
\be
 P^{(\infty)} \approx e^{-t^2/N\tau_Z^2} \ \longrightarrow \ 1 .
 \label{zeno6}
\ee
That is, in the ideal limit where the system can be monitored
indefinitely, its quantum state will remain the same, without
evolving, because the interaction is not enough to remove its
quantum coherence.
In this way, by means of consecutive measurements, the system decay is
slowed down. This is the QZE, predicted
by Misra and Sudarshan in 1977 for an unstable particle
\cite{misra}. Later on, in 1989, Itano {\it et al.}\ \cite{itano} observed
how Be atoms were inhibited to evolve into the excited state by
means of ultraviolet pulses. More recently, Fischer {\it et al.}\
\cite{fischer} observed QZE with cold Na atoms in an
experiment trying to mimic the proposal of Misra and Sudarshan.

In the experiments conducted by Fischer {\it et al.}\
\cite{fischer}, AZE was also found, which was predicted by Kofman
and Kurizki \cite{kofman,kurizki} within the density matrix
formalism. This phenomenon occurs when the decay process is
accelerated due to continuous measurements (ideal and sufficiently
frequent). Notice that the survival probability (\ref{zeno5}) can
also be expressed as a general function of the measurement time,
$\tau$, as \cite{facchi,pascazio}
\be
 P^{(N)}(t) = e^{-\gamma_{\rm eff}(\tau) t} ,
 \label{zeno7}
\ee
%
%
%
If $\tau$ is small compared to $\tau_Z$ (although not always this
constitutes an appropriate time scale \cite{pascazio}), then
$\gamma_{\rm eff} \to \tau/\tau_Z^2$, as in (\ref{zeno6}).
However, for $\tau$ large enough (but still small), the system
should display a typical decay at a constant rate $\gamma_{\rm
free}$, as it is found in unstable systems according to Fermi's
golden rule, i.e., $\gamma_{\rm eff} \to \gamma_{\rm free}$. The
accelerated decay with respect to the exponential decaying
behavior is the AZE. The general decay law ruling the behavior of
unstable quantum systems has been discussed in detail by Peres
\cite{peres1}. The physical conditions for leading to the
observation of QZE and AZE have been established in a very
illuminating way by Kofman and Kurizki \cite{kurizki}. If the
spectral density of states coupled to the initial state is a dense
band or continuum (acting as a reservoir), the
measurement-modified decay rate can be expressed as a simple
overlap between the spectral density of final states or reservoir
coupling spectrum, $G(\omega)$, and the measurement-induced
initial state level width, $F(\omega;\tau)$, according to
\begin{equation}
 \gamma_{\rm eff} (\tau) = 2 \pi \int_0^\infty G(\omega)
 F(\omega;\tau) d \omega ,
 \label{gamma}
\end{equation}
where the measurement frequency $\bar{\lambda} = \tau^{-1}$ (we
use the bar over $\lambda$ in order to avoid confusions with the
collisional frequency from surface diffusion), is related to the
initial state energy uncertainty, $\Delta E$, as $\Delta
E/\bar{\lambda} \sim \hbar$. Hence the decay rate is essentially
determined by the spectral density profile within a bandwidth
around its energy level. According to the same authors
\cite{kurizki}, (\ref{gamma}) constitutes a universal result:
frequent measurements on a given initial state generally lead to
its dephasing through randomization of the corresponding phase.
The broadening of the initial state is of the order of the
measurement frequency and it can be seen as an analog of the
collisional broadening leading to a phase randomization of the
state. Two extreme cases are then envisaged. On the one hand, when
the measurement frequency $\bar{\lambda}$ is much greater than the
spectral density width and the detuning between the reservoir
center of gravity and the initial state frequency position, the
QZE holds (the decay rate goes like $\bar{\lambda}^{-1}$). A
reduction of the decay rate is then obtained when compared to the
measurement-free (or Fermi's golden rule) decay. Mathematically,
the spectral density is assumed to be a Dirac $\delta$ peak. In other
words, when an Ohmic friction or white noise, characterized by a
linear spectral density, is assumed, the QZE will not be
observable. Non-Ohmic reservoirs should be then considered with a
cut-off frequency. On the other hand, if the measurement frequency
is much smaller than the corresponding detuning, the decay rate is
shown to grow with $\bar{\lambda}$, which leads to AZE. Actually,
the AZE seems to be much more ubiquitous than the QZE
\cite{kurizki}. More recently, Maniscalco {\it et al}
\cite{piilo}, within the density matrix formalism, have
established the conditions for the occurrence of such effects
within the quantum Brownian motion; in particular, they have
studied a quantum harmonic oscillator linearly coupled to a
quantum reservoir modeled as a collection of non-interacting
harmonic oscillators at thermal equilibrium. During the
time-evolution, the system was subject to a series of
non-selective measurements, i.e., measurements which do not
select the different outcomes \cite{petruccione}. The
factorization displayed in (\ref{zeno5}) follows from the fact
that at second order in the coupling, the density matrices of the
system and environment factorize at any time \cite{lax}.

Within the scenario of the surface diffusion of interacting adsorbates,
and according to (\ref{gamma}), the collisional friction $\lambda$
would govern the $F$ function, and the surface friction $\gamma$,
the $G$ function.
Furthermore, it is interesting to note the following comparison.
Within the standard projective-measurement scenario described above,
a series of $N$ measurements is carried out at regular intervals of
time; within the non-selective-measurement scenario, $\lambda$
provides an average number of collisions per time unit \cite{ruth6}
and, therefore, an average number of measures per time unit, although
such measures are carried out at random (according to a Poissonian
distribution).


\section{Surface diffusion of a wave packet by a flat surface}
\label{sec2}

Consider the diffusion of a wave packet through a flat surface. It
is the simplest integrable dissipative quantum system we may
devise. This simple example is quite illustrative to show the
different conclusions about the appearance of the QZE and AZE when
carrying out ideal (or projections) and indirect measurements.


\subsection{High surface temperatures}
\label{sec21}

Within the ISA model, the quantum Langevin equations for the
Heisenberg position operators read (for Ohmic friction or
linear spectral density) as
\be
 \begin{array}{l}
 \ddot{x}(t) = - \eta \dot{x}(t) + \delta F_x (t) , \\
 \ddot{y}(t) = - \eta \dot{ y}(t) + \delta F_y (t) ,
 \end{array}
 \label{eq6}
\ee
where ${\bf R} = (x,y)$ denotes the position operators of a single
adsorbate moving on a flat surface and the ``dots'' over the position
operators denote time-derivatives. In this model, two
non-correlated noises (per mass unit) are assumed to simulate the
two baths: a Gaussian white noise, accounting for the lattice
vibrational effects due to the surface temperature and leading to
the interaction with the adsorbates, and a white shot noise, which
simulates the adsorbate-adsorbate collisions. Then, for each
degree of freedom, we have $\delta F_i (t) = \delta F_{i,G}(t) +
\delta F_{i,S} (t)$ where $i=x,y$, the noise fluctuation is given
by $\delta F = F - \langle F \rangle$, and $G$ and $S$ stand for
the Gaussian and shot noise, respectively. At high surface
temperatures, $\beta^{-1} \gg \hbar \eta$ (or $\eta^{-1} \gg t_c
\sim \hbar \beta$), noise is considered to be classical and its
autocorrelation function at two different times is well described
by a Dirac $\delta$-function, as assumed in (\ref{eq6}). In other
words, the time interval between these two times has to be greater
than $t_c$.

Consider an adparticle of mass $m$ initially placed on a given
position of a flat surface and represented by a Gaussian state
\begin{equation}
 \psi (x,y,0) = \frac{1}{\sqrt{2\pi\sigma_0^2}}\
  e^{- x^2/4\sigma_0^2 - y^2/4\sigma_0^2} ,
 \label{g0}
\end{equation}
where the initial width along each direction is the same and equal
to $\sigma_0$. The adparticle is assumed to be initially in
equilibrium with the reservoir or heat bath (surface) at a
temperature $T$, but weakly coupled to the environment that
dissipation can be neglected. The role of the initial conditions
has been very often discussed in the literature (see, for example,
\cite{hakim,Ford}). After a time $t$, the probability to find the
particle at a given position $(x,y)$ is given by averaging the
survival probability over a thermal (Maxwell-Boltzmann)
distribution of velocities
\cite{hakim,Ford,lewis,ford1,ford2,dorlas}
%
%
\begin{equation}
 P(x,y,t) = \frac{1}{2\pi w_x(t) w_y(t)}\
  e^{- x^2/2w_x^2(t) - y^2/2w_y^2(t)} .
 \label{pt}
\end{equation}
Note here that the interaction with the environment makes the
quantum state describing the system to pass from pure to a
statistical mixture. Therefore, the probability (\ref{pt}) has to
be interpreted as a conditional probability, rather than the
probability density associated with a pure state. According to
Ford {\it et al.}\ \cite{ford1,ford2}, this normal distribution is
associated with two measurements at two different times. For each
degree of freedom, the overall time-dependent spreading of the
wave packet can be written as
\begin{equation}
w_i^2(t) = \sigma_0^2 + \sigma_i^2(t) + s_i(t) ,
\label{wt}
\end{equation}
with $i=x,y$. Here the quantum contribution to the spreading is
given by
\be
 \begin{array}{l}
  \sigma_x^2(t) = \displaystyle - \frac{[x(0),x(t)]^2}{4\sigma_0^2} , \\
  \sigma_y^2(t) = \displaystyle - \frac{[y(0),y(t)]^2}{4\sigma_0^2} ,
 \end{array}
 \label{sigq}
\ee
and $s_i(t)$ is the mean-square displacement (MSD) along each
direction,
\be
 \begin{array}{l}
  s_x(t) = \langle \{ x(t)-x(0) \}^2 \rangle , \\
  s_y(t) = \langle \{ y(t)-y(0) \}^2 \rangle .
 \end{array}
 \label{st}
\ee

In order to understand the effects of the measurements on the quantum
system, apart from the survival probability seen in (\ref{sec0}), one can
also analyze the decay rate of its initial state by monitoring the ratio
between the probabilities at $t$ and $t=0$, and  evaluated at $(x,y) =
(0,0)$ \cite{dorlas}.
Thus, in the case of the initial Gaussian state (\ref{g0}), we find
\begin{equation}
 R(t) = \frac{\sigma_0^2}{w_x(t)\ w_y(t)} \label{rt} ,
\end{equation}
which, taking into account (\ref{wt}), can be explicitly written as
\begin{equation}
 R(t) = \frac{\sigma_0^2}{w^2(t)} = \frac{\sigma_0^2}{\sigma_0^2 +
  s(t) + \sigma^2(t)} ,
 \label{Rt}
\end{equation}
since the spreading along each direction is obviously the same.
Accordingly, this calculation reduces to the simple evaluation of the
quantum spreading $\sigma_i(t)$ and the MSD $s_i(t)$ along each
direction ($i=x,y$).
The same time-dependence is obtained if Gaussian integrations are
carried out in the corresponding ratio instead of evaluating it at
$(x,y) = (0,0)$. The quantum spreading depends on the commutator
of the position operators at two different times, from which
\begin{equation}
 \sigma_i^2(t) = \frac{\hbar^2}{4 m^2 \sigma_0^2 \eta^2}\
  \Phi^2 (\eta t) ,
 \label{qs}
\end{equation}
with $\Phi (\eta t) = 1 - \exp(-\eta t)$.
For each degree of freedom, one obtains the same spreading because the
Gaussian initial with is assumed to be the same for both directions.
Analogously, the MSD along each direction takes the form
\begin{equation}
s_i(t) = \frac{2 \hbar \eta}{\pi m}\ t^2 H(\eta t;T) .
\label{sint}
\end{equation}
Here, the function
\begin{equation}
 H(\eta t;T) = \int_0^{\infty}
  \frac{1 - \cos z}{z (z^2 + \eta^2 t^2)}\
  {\rm coth} \left( \frac{\hbar z}{2 t k_B T} \right) dz
\label{int}
\end{equation}
has an analytical solution, which allows us to express (\ref{sint}) as
\begin{equation}
 s_i(t)  =  \frac{2 k_B T}{m \eta} \left[ t - \frac{1}{\eta}
  \Phi(\eta t) \right] .
 \label{sint-f1}
\end{equation}

In order to analyze (\ref{Rt}) now for high surface temperatures, we
will consider two time regimes: $\eta t \ll 1$ (short time) and
$\eta t \gg 1$ (long time).
In the short-time regime, $\eta t \ll 1$, the quantum spreading
(\ref{qs}) goes like $t^2$, according to
\begin{equation}
\sigma^2 (t) \approx \frac{\hbar^2}{4 m^2 \sigma_0^2}\ t^2 ,
 \label{qs-lowt}
\end{equation}
which corresponds to the wave packet spreading in the absence of
dissipation. Analogously, the MSD also goes like $t^2$ according
to
\begin{equation}
s(t) \approx \frac{k_B T}{m}\ t^2 , \label{shighT-lt}
\end{equation}
where the prefactor is the thermal velocity in two dimensions.
Thus, the overall time-dependent spreading can be expressed in a
more compact form as
\begin{equation}
 w^2(t) \approx \sigma_0^2 + \langle v^2 \rangle t^2 \label{wtt} ,
\end{equation}
with
\begin{equation}
 \langle v^2 \rangle = \frac{k_B T}{m} +
 \frac{\hbar^2}{4 m^2 \sigma_0^2} .
 \label{v2}
\end{equation}
In this short-time regime (friction-free motion or ballistic
regime), we have usually that $\langle v^2 \rangle t^2 \ll
\sigma_0^2$ (the wave packet has not spread too much compared to
its initial spreading which can be assumed to be arbitrary large)
and therefore (\ref{Rt}) becomes
\begin{equation}
 R(t) \approx 1 - \frac{\langle v^2 \rangle}{\sigma_0^2} t^2 + \ldots
 \approx e^{- \langle v^2 \rangle t^2 / \sigma_0^2} ,
 \label{shortR}
\end{equation}
which is the standard short-time, $t^2$-behavior usually associated
with QZE. Notice that the effective decay rate ($\eta_{\rm eff} (\tau)
= \tau \sqrt{\langle v^2 \rangle} / \sigma_0$) is independent of
the total friction $\eta$ and, in particular, of $\lambda$. Hence
the dynamical system displays a friction-free motion with a
quadratic time behavior of the decay rate. By replacing
$\tau$ by $t/N$ (with $N \rightarrow \infty$), some authors
\cite{dorlas} show that the factorization given in
(\ref{zeno5}) also applies for $R(t)$ and claim that in the
absence of friction, the QZE always holds. However, after Kofman
and Kurizki, for Ohmic friction or linear spectral density, the
QZE is not expected to occur (see Section~\ref{sec0}). In the
indirect measurement scheme used here, the dynamical system and
the reservoir are entangled at all times except for such a regime,
since $\eta = \gamma + \lambda$ is not playing any role yet. Thus,
the QZE does not hold because any indirect measurement through
$\lambda$ has been carried out. Moreover, decoherence is absent in
this short-time regime since it is a free-motion regime.

In the long-time (or diffusion) regime, $\eta t \gg 1$, the quantum
spreading in both directions is time-independent, with
\begin{equation}
 \sigma^2 (t \rightarrow \infty)  \approx \frac{\hbar^2}{4 m^2
  \sigma_0^2 \eta^2} .
 \label{qs-hight}
\end{equation}
On the contrary, the corresponding MSD is linear with time,
according to
\begin{equation}
s(t) \approx \frac{2 k_B T}{m \eta}\ t
 \label{shighT-gt} ,
\end{equation}
and Einstein's law is fulfilled, with the diffusion constant given by
\begin{equation}
 D = \frac{k_B T}{m \eta}
 \label{D}
\end{equation}
in both directions for an isotropic surface.
Thus, the full time-dependent spreading can be expressed as
\begin{equation}
 w^2(t) \approx \sigma_0^2 + \frac{\hbar^2}{4 m^2 \sigma_0^2
 \eta^2} + \frac{2 k_B T}{m \eta}\ t  .
 \label{wtl}
\end{equation}
Assuming $\sigma_0^2 \gg s(t) \gg \sigma^2(t)$ for a certain
time within the long-time regime, $R(t)$ can be approximated by
\begin{equation}
 R(t) \approx 1 - \frac{2 k_B T}{m \sigma_0^2 \eta}\ t .
 \label{largeR}
\end{equation}
Therefore, by increasing the coverage or collisional frequency,
$\lambda$, according to (\ref{largeR}), the decay rate decreases,
but obviously one cannot speak about the QZE, because the system
has already relaxed dynamically and decoherence is already
manifested.


\subsection{Low and zero surface temperatures}
\label{sec22}

At low surface temperatures ($\eta^{-1} \ll t_c \sim \hbar
\beta$), the noise autocorrelation function is complex
\cite{ingold} and depends on the ratio between the interval of the
two times and $t_c$ (colored noise). In general, the
noise function acts like a driving force and the surface dynamics
is better described within the generalized Langevin framework as
\be
 \begin{array}{l}
  \ddot{x}(t) +
   \displaystyle \int_{-\infty}^t dt' \eta_x(t-t') \dot{x}(t')
   = \delta F_x (t) , \\
  \ddot{y}(t) + \displaystyle \int_{-\infty}^t dt' \eta_y(t-t')
   \dot{y}(t)=\delta F_y (t) ,
 \end{array}
 \label{xy-gl}
\ee
where $\eta_x(t)$ and $\eta_y(t)$ represent the time-dependent
frictions or memory functions along each direction. The
one-dimensional expression for the noise function can be found in
the literature \cite{lewis}. If we assume the surface is isotropic
and both frictions are Ohmic, (\ref{xy-gl}) reduces to
(\ref{eq6}). Thus, the corresponding quantum mechanical process is
not a Markovian process in the customary sense of the term
\cite{lewis}. The quantum spreading is the same as before,
i.e., given by (\ref{qs}), since it is independent of the surface
temperature. However, the MSD along each direction, given by
(\ref{sint}), now reads as
\be
\fl
 s_i(t) = \frac{2 k_B T}{m \eta} \left[ t - \frac{1}{\eta}
  \Phi(\eta t) \right]
  + \frac{4}{\beta m} \sum_{n=1}^{\infty} \frac{\eta - \nu_n -
  \eta e^{- \nu_n t} + \nu_n e^{-\eta t}} {\nu_n (\eta^2 - \nu_n^2)} ,
 \label{sint-f11}
\ee
with
\begin{equation}
\nu_n = \frac{2 \pi n}{\hbar \beta}
 \label{mat}
\end{equation}
being the so-called {\it Matsubara frequencies}, which come from
the Taylor series expansion of the coth-function in the integral
of (\ref{int}). At high surface temperatures, (\ref{sint-f11})
reduces to (\ref{sint-f1}). The sum over $n$ plays an important
role only at very low surface temperatures. Nevertheless, at zero
surface temperature, the sum disappears, since the coth-function
becomes unity and $H(\eta t; 0)$ acquires a different dependence
on time. In this case ($T=0$), the corresponding MSD expression
reads as
\begin{equation}
 s_i(t) = \frac{2 \hbar}{\pi m \eta} \left \{ \gamma_E + \ln \eta t
  - \frac{1}{2} \left[ e^{\eta t} {\bar E}i(- \eta t)  + e^{- \eta t}
  Ei(\eta t) \right] \right \} ,
 \label{sint-f2}
\end{equation}
where $\gamma_E=0.577$ is Euler's constant, and ${\bar E}i(- \eta
t)$ and $Ei(\eta t)$ are the exponential integrals
\cite{grandshteyn}. The environment no longer transfers energy to
the adparticle due to the zero-point motion of the surface
oscillators.

A similar dynamical analysis can also be carried out in terms of
the short- and long-time regimes. The quantum spreading in both
dimensions remains the same and is given by (\ref{qs-lowt}) and
(\ref{qs-hight}), respectively. However, the MSD is different due
to its temperature dependence. Thus, in the short-time regime,
$\eta t \ll 1$, we find
\begin{equation}
 s(t) \approx \frac{k_B T}{m}\ t^2 + \frac{\hbar \eta}{\pi m}\ t^2
 \label{st-s}
\end{equation}
which depends linearly on $\eta$. Analogously, at zero surface
temperature, we have
\begin{equation}
 s(t) \approx \frac{\hbar \eta}{\pi m}\ t^2\
  \left(\frac{3}{2} - \gamma_E - \ln \eta t \right) ,
 \label{st-s0}
\end{equation}
where we observe essentially the same friction and time
dependence, since the presence of the friction in the log-function
is much weaker. On the other hand, in the long-time regime, it can
be shown that
\begin{equation}
 s(t) \approx \frac{2 k_B T}{m \eta}\ t + \frac{2 \hbar}{\pi m}
  \frac{1}{\eta+\nu_1} ,
 \label{stl-l}
\end{equation}
which, at zero surface temperature, becomes
\begin{equation}
 s(t) \approx \frac{2 \hbar}{\pi m \eta}
  \left( \gamma_E + \ln \eta t \right).
 \label{st-l0}
\end{equation}
Taking into account these behaviors, the following conclusions are
drawn at low and zero surface temperatures:
\begin{itemize}

\item Short-time regimes. The same formal expression as that given by
(\ref{shortR}) is obtained for $R(t)$, except for now $\langle v^2
\rangle$ depends linearly on $\eta$, i.e.\ on $\lambda$, even
at zero temperature. Thus, by increasing the coverage or
$\lambda$, $R(t)$ decreases (contrary to the manifestation of the
QZE).

\item Long-time regime. The $R(t)$ function can be expressed as
\begin{equation}
 R(t) \approx 1 - \frac{s(t)}{\sigma_0^2} ,
 \label{large-R}
\end{equation}
where the MSD goes with $\lambda^{-1}$ according to (\ref{stl-l})
and (\ref{st-l0}). Thus, by increasing $\lambda$, the decay rate
decreases. However, as before, this manifestation
cannot be attributed to the QZE, because the relaxation process
has already been established and decoherence dominates the
diffusion process. The AZE is not found either.
\end{itemize}

For ideal measurements, at zero surface temperatures, Dorlas and
O'Connell \cite{dorlas} showed that the QZE is characterized by
small $\eta t$ values, whereas the AZE by large values of $\eta t$.
According to the previous analysis, though, we conclude again
differently due to the non-selectiveness or indirectness of the
measurements we have considered on the time evolution of the
adparticle.


\section{The quantum intermediate scattering function}
\label{sec3}


\subsection{Flat surfaces}

For flat or weakly corrugated surfaces where the thermal energy,
$k_B T$, is greater than the diffusion barrier, the quantum motion
of the adparticles can be exactly described within the Heisenberg
picture and the scenario of the ISA model \cite{ruth3,ruth2}. As
mentioned before, this study can be carried out by means of the
so-called intermediate scattering function. This function provides
us information about the time decay associated with the position
of the adparticle, which is initially located at ${\bf R} (0)$
and, after a time $t$, it will be in a position ${\bf R} (t)$, as
it can be inferred from the autocorrelation function
(\ref{zeno10}).
%
%
In spin-echo experimental techniques, this function is
proportional to the polarization, with the real and imaginary
parts of the latter being observable magnitudes
\cite{peter,allison1,allison2}.

In general, an exact, direct calculation of $I(\Delta {\bf K},t)$
results difficult to carry out due to the non-commutativity of
the adparticle position operators at different times.
The product of the two exponential operators in (\ref{zeno10}) can
be evaluated according to a special case of the Baker-Hausdorff
theorem ({\it disentangling theorem}), namely $e^{\hat{A}}
e^{\hat{B}} = e^{[\hat{A}, \hat{B}]/2} e^{\hat{A}+\hat{B}}$, with
$\hat{A} = - i \Delta {\bf K} \cdot {\bf R}(0)$ and $\hat{B} = i
\Delta {\bf K} \cdot {\bf R}(t)$, which holds whenever the
corresponding commutator is a c-number. When no adsorbate-substrate
interaction potential is assumed, the formal solution of
(\ref{eq6}) can be written as \cite{vineyard,ruth3,ruth2}
\begin{equation}
 {\bf R} (t) =  {\bf R} (0) + \frac{{\bf P} (0)}
 {m \eta} \ \! \Phi (\eta t) +  \frac{1}{m \eta} \int_0^t
 \Phi (\eta t - \eta t') \delta {\bf F} (t') dt' ,
 \label{eq5}
\end{equation}
where ${\bf P} (0)$ is the initial adparticle momentum operator.
The commutator of the position operators at two different times is
a $c$-number, since the commutator $[{\bf R}(0),\delta {\bf F}]$
is zero. This can be justified considering the Caldeira-Leggett
model, where the two noise functions depend on time, on the
initial positions and momenta of the harmonic oscillators
associated with the two baths, and therefore only on the
adparticle initial position \cite{salva1,ruth1,ruth2}. Thus,
$I(\Delta {\bf K},t)$ can be expressed as
\begin{equation}
 I (\Delta {\bf K},t) =
  I_q (\Delta {\bf K},t) I_c (\Delta {\bf K},t) ,
 \label{isf2}
\end{equation}
i.e., as the product of a quantum-mechanical intermediate
scattering function, $I_q$, and a classical-like one, $I_c$.

Assuming a Markovian scenario, $I_q$ reads as \cite{ruth3,ruth2}
\begin{equation}
 I_q (\Delta {\bf K},t)
  = \exp \left[ \frac{i E_r}{\hbar \eta} \Phi(\eta t) \right] ,
 \label{iq}
\end{equation}
where $E_r = \hbar^2 \Delta {\bf K}^2/2m$ is the {\it adsorbate
recoil energy}, which becomes less important as the adparticle
mass increases. The $I_q$-factor is a time-dependent phase with
its argument decreasing with the total friction and therefore with
the coverage; its time-dependence only comes through the
$\Phi(\eta t)$ function. At short-times ($\eta t \ll 1$ or $\hbar
\beta \ll \eta^{-1}$), $\Phi(\eta t) \approx \eta t$ and the
argument of $I_q$ becomes independent of the total friction, thus
increasing linearly with time. On the contrary, in the asymptotic
time limit ($\eta t \gg 1$ or $\hbar \beta \gg \eta^{-1})$, this
argument approaches a constant phase. Regarding the classical-like
factor, $I_c$, in the Gaussian approximation, it takes the form
\cite{ruth2}
\begin{equation}
I_c(\Delta {\bf K},t) \simeq  e^{- \Delta {\bf K}^2 \int_0^t
(t-t') C_v(t') dt'} = e^{- \Delta {\bf K}^2 [f(t) + g(t)]} ,
 \label{ic}
\end{equation}
where $C_v(t)$ is the velocity autocorrelation function, which can
be expressed in terms of the time-dependent functions $f(t)$ and
$g(t)$, given by
\begin{equation}
f(t) = \left( \frac{1}{m \beta \eta^2} - i \frac{\hbar}{2 m
\eta}
 \right) [ e^{- \eta t} + \eta t -1]
 \label{ft}
\end{equation}
and
\begin{equation}
g(t) = \frac{2}{m \beta} \sum_{n=1}^{\infty} \frac{\nu_n e^{-
\eta t} - \eta e^{- \nu_n t} + \eta - \nu_n} {\nu_n (\eta^2 -
\nu_n^2)} .
 \label{gt}
\end{equation}
The total intermediate scattering function (\ref{isf2}) can
then be expressed as
\begin{equation}
I(\Delta {\bf K},t) = e^{- \chi^2 [\alpha^* \eta t - \Phi (\eta
t)]} e^{- \Delta {\bf K}^2 g(t)} ,
 \label{isf3}
\end{equation}
with $\chi^2 = \Delta {\bf K}^2 \langle v^2 \rangle / \eta^2$ and
$\alpha = 1 + i \hbar \beta \eta / 2$, $\langle v^2 \rangle = 1/m
\beta$ being the thermal square velocity. The recoil energy is
included in the imaginary part of the product $\chi^2 \alpha^*$,
which disappears when $\hbar \rightarrow 0$. Equation~(\ref{isf3})
is the generalization of the intermediate scattering function for
the quantum motion of interacting adsorbates in a flat surface.
The dependence of this function on $\Delta {\bf K}^2$ through the
so-called shape parameter, $\chi$, is the same as in the classical
theory \cite{ruth3}. From now on, we will just focus on
(\ref{isf3}), which will be analyzed in terms of the QZE and AZE
considering the same two time regimes as before (short- and
long-times).

In the short-time (ballistic or free-diffusion) regime ($\eta t \ll 1$),
(\ref{isf3}) becomes
\begin{equation}
I(\Delta {\bf K},t) \approx e^{i E_r t / \hbar}  ,
 \label{isf-lt}
\end{equation}
where the real part has also the typical $t^2$-behavior
characteristic of the QZE,
\begin{equation}
{\rm Re} \{ I(\Delta {\bf K},t) \} \approx 1 - \frac{1}{2}
\frac{E_r^2}{\hbar^2} t^2 ,
 \label{re-lt}
\end{equation}
which is friction free. As before, not always one has a quadratic
time dependence, the QZE holds. Indeed, with Ohmic environments,
as in our case, the QZE does not take place. The imaginary part,
in turn, is linear with time, according to
\begin{equation}
{\rm Im} \{ I(\Delta {\bf K},t) \} \approx \frac{E_r}{\hbar} t .
 \label{im-lt}
\end{equation}
The short-time regime is related to a Gaussian behavior of the
real part of the intermediate scattering function, which has been
observed in real experiments \cite{allison1,allison2}. The time
Fourier transform of this function is the so-called dynamic
structure factor (which is also observable), leading to Gaussian
wings in the corresponding quasi-elastic peak. This frictionless
motion (no collisions among adsorbates and no interaction with the
substrate) is the regime where dynamical coherence prevails,
i.e., adparticles keep their memory of velocity.

In the long-time (diffusive) regime ($\eta t \gg 1$),
(\ref{isf3}) can be written as
\begin{equation}
I(\Delta {\bf K},t) \approx e^{- \chi^2 \eta t} e^{i E_r t /
\hbar} ,
 \label{isf-gt}
\end{equation}
where it is clear that the real part exponentially annihilates any
trace of the oscillations associated with the phase depending on the
recoil energy.
Here, the Markovian character of the evolution is very
noticeable and the regime of coherent dynamics is completely lost,
which is a typical trait of diffusion regimes. The corresponding
dynamic structure factor displays a Lorentzian shape around zero
energy transfers of the quasi-elastic peak. This gradual change of
shape in the dynamic structure factor is known as the {\it
motional narrowing} effect \cite{vega1,vega2,ruth4}. Actually,
strictly speaking, the quasi-elastic peak is a mixture of both
shapes: Lorentzian-like at very small frequencies and
Gaussian-like at very large frequencies. According to
(\ref{isf-gt}), the real part of the intermediate scattering
function decays slower with the coverage and therefore the dynamic
structure factor becomes narrower (Gaussian wings)
\cite{ruth6,ruth5}. At this time regime, decoherence is already
manifested (or relaxation process). In other words, the loss of
coherence (or loss of quantum features) begins at times greater
than the coherence time $t_c$. Quantum motion only plays an
important role around this time and when $t \gg t_c$, the
classical intermediate scattering function dominates, since the
commutator in $I_q$ plays no role and $I_c$ approaches the
classical result. The lowest surface temperature reached in a
typical $^3$He spin echo experiment is around 100~K, which means
$t_c = 0.07$4~ps. For neutron spin echo measurements, the surface
temperature can be slowed down up to a few Kelvin, leading to a
better analysis for quantum coherence.

Quasi-elastic He-atom scattering measurements \cite{ellis}
constitute an interesting example worth mentioning, since they
provide a clear evidence for a two-dimensional free gas of Xe
atoms on Pt(111) at low coverages ($\theta = 0.017$), low incident
atom energy ($E_i = 10.15$~meV) and surface temperature of 105~K
along the (100) direction. The corrugation of the surface is
assumed to be negligible at that surface temperature, which
manifests in a dynamic structure factor displaying a Gaussian-like
line shape.
Apart from quite different mean free paths extracted from
the experiment \cite{ellis} and from our own analysis
\cite{ruth4}, one would be tempted to consider this system an
ideal candidate to observe QZE, since the intermediate scattering
function is quadratic in time for a quite long time interval.
However, due to the heavy adsorbate mass, the recoil energy is
expected to be really small and a classical diffusion motion
should be good enough to describe this two-dimensional free gas.
The classical intermediate scattering function for a flat surface
is
\begin{equation}
 I_{\rm classical} (\Delta K,t) = \exp \left[- \chi^2
   \left( e^{- \eta t} + \eta t - 1 \right) \right] ,
 \label{eq:isf4}
\end{equation}
which comes from considering $\hbar = 0$ in the complex number
$\alpha$ and $g(t) =0$ in (\ref{isf3}). When $\eta t \ll 1$,
the $t^2$-behavior is again obtained
\begin{equation}
 I_{\rm classical} (\Delta K,t) \approx \exp ( - \delta^2 t^2 ) ,
 \label{eq:isf5}
\end{equation}
with $\delta = \Delta {\bf K}^2 k_B T / 2 m$. This behavior with
time remains until the adparticle has travelled a distance of the
order of the mean free path and it starts to lose its coherence.
But again, for Ohmic friction, the QZE does not hold.


\subsection{Corrugated surfaces}
\label{sec31}

Surface diffusion of adsorbates on  corrugated surfaces can be
very often described in the Markovian approximation and ISA model
by the standard Langevin equation \cite{ruth2}
\begin{equation}
 m \ddot{\bf R} = - m \eta \dot{\bf R} + \nabla V({\bf R})
  + \delta {\bf F} ,
 \label{LME}
\end{equation}
where $V({\bf R})$ is the adiabatic interaction potential between
the adsorbate and the surface. Again the formal solution of
(\ref{LME}) is given by
\be
\fl
 {\bf R} (t) = {\bf R} (0) + \frac{{\bf P} (0)}{m \eta} \ \! \Phi (\eta t)
  + \frac{1}{m \eta} \int_0^t
   \Phi (\eta t - \eta t') [\nabla V ({\bf R}(t')) + \delta
   {\bf F} (t')] dt' .
 \label{LME-s}
\ee
The presence of the adiabatic force introduces an additional
commutator, $[{\bf R}_0,\nabla V({\bf R}(t))] = (i \hbar)\partial
\nabla V({\bf R}(t))/\partial {\bf P}_0$, where the dependence of
this force on the initial state $({\bf R} (0), {\bf P} (0))$ is
through ${\bf R}(t)$.  Within a quantum Markovian framework
\cite{ruth3}, this commutator is very small or negligible. Thus,
in the quantum intermediate scattering function
(\ref{isf2}), the $I_q$ factor is the same as in the case of a
flat surface, given by (\ref{iq}), determining again the
very short time, friction free behavior of the real part of the
total intermediate scattering function.
The classical-like factor, $I_c$, deserves special attention, since
exact quantum Langevin calculations for corrugated surfaces are in
general, prohibitive. Only the long time regime can be analyzed in
simple terms. Quantum corrections at different coverages and
temperatures in the activated surface diffusion of Na atoms on
Cu(001) have been analyzed \cite{ruth3}. We have shown that within
the Gaussian approximation, Eq. (\ref{ic}), and by assuming a
velocity autocorrelation function given by
\begin{equation}
 C_v(t) = \frac{k_B T}{m} \ \! e^{- \tilde{\eta} t}
  \cos (\tilde{\omega}t + \tilde{\delta}) ,
 \label{corrg}
\end{equation}
$I_c$ becomes
\begin{eqnarray}
 I_c(\Delta K, t) & = & e^{-\chi_l^2 \tilde{f}(\tilde{\omega},t)}
 \nonumber \\
 & = & e^{-\chi_l^2 \tilde{A}_1 - \chi_l^2 \tilde{A}_2 t}
 \nonumber \\
 & \times & \!\!\! \sum_{m,n=0}^\infty \frac{(-1)^{m+n}}{m! \ \! n!} \ \!
  \chi_l^{2(m+n)} \tilde{A}_3^m \tilde{A}_4^n
  e^{-(m+n)\tilde{\eta}t + i(m-n)(\tilde{\omega}t + \tilde{\delta})} ,
  \nonumber \\
 \label{isfg}
\end{eqnarray}
where
\be
 \tilde{f}(\tilde{\omega},t) \equiv \tilde{A}_1 + \tilde{A}_2 t
  + \tilde{A}_3 e^{i\tilde{\delta}} e^{-(\tilde{\eta}
   - i\tilde{\omega})t}
  + \tilde{A}_4 e^{-i\tilde{\delta}} e^{-(\tilde{\eta}
   + i\tilde{\omega})t} ,
\ee
and
\ba
 \tilde{A}_1 & = & \frac{\tilde{\eta}^2
  [2 \tilde{\eta} \tilde{\omega} \sin \tilde{\delta}
  + (\tilde{\omega}^2 - \tilde{\eta}^2) \cos \tilde{\delta})}
  {(\tilde{\eta}^2 + \tilde{\omega}^2)^2} ,
 \\
 \tilde{A}_2 & = &
  \frac{\tilde{\eta}^2 (\tilde{\eta} \cos \tilde{\delta}
   - \tilde{\omega} \sin \tilde{\delta})}
  {\tilde{\eta}^2 + \tilde{\omega}^2} ,
 \\
 \tilde{A}_3 & = & \frac{\tilde{\eta}^2}
  {2(\tilde{\eta} - i\tilde{\omega})^2} ,
 \\
 \tilde{A}_4 & = & \frac{\tilde{\eta}^2}
  {2(\tilde{\eta} + i\tilde{\omega})^2} .
\ea
Usually, the values of the parameters $\tilde{\eta}$,
$\tilde{\omega}$ and $\tilde{\delta}$ are obtained from a fitting
of the numerical results issued from solving the standard Langevin
equation with periodic boundary conditions to (\ref{corrg}).
Moreover, $\chi_l (\Delta {\bf K})$ is a generalized shape
parameter proposed by the authors to be \cite{jcp-elliot}
\be
 \chi_l (\Delta {\bf K}) \equiv \sqrt{\frac{\Gamma_{\nu}
 (\Delta {\bf K})}{2 \eta}} ,
 \label{eq11b}
\ee
where $\Gamma_{\nu} (\Delta {\bf K})$ represents the inverse of
the correlation time and is expressed as
\be
 \Gamma_{\nu} (\Delta {\bf K}) = \nu
 \sum_{\bf j} P_{\bf j} \ \! [1 - \cos({\bf j} \cdot \Delta {\bf K})] ,
 \label{eq9b}
\ee
with $\nu$ being the total jump rate out of an adsorption site and
$P_{\bf j}$ the relative probability that a jump with a
displacement vector ${\bf j}$ occurs.

\begin{figure}
 \begin{center}
 \epsfxsize=10cm {\epsfbox{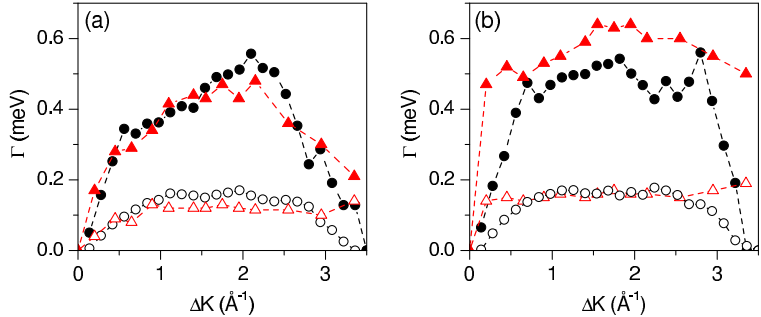}}
 \caption{\label{fig1}
  Diffusion of Na atoms on Cu(001). Numerical (triangles) and
  experimental (circles) dependence of the broadening of the Q-peak
  on $\Delta K$, at 200~K (open circles and triangles) and 300~K (full
  circles and triangles) along the azimuth [100] for two different
  values of the surface coverage: (a) $\theta = 0.064$ and
  (b) $\theta = 0.106$.}
 \end{center}
\end{figure}

Notice the linear dependence on time in $\tilde{f}$ because of the
parameters $\tilde{\eta}$, $\tilde{\omega}$ and $\tilde{\delta}$
are time-independent. This leads to an effective decay rate in
(\ref{isfg}) given by $\chi_l^2 \tilde{A}_2$, which accounts for
the diffusion and causes the intermediate scattering function
to vanish at asymptotic times. This scattering function also
provides information on the low vibrational motion or T-mode. This
fact is better appreciated in the dynamic structure factor which
is the time Fourier transform of the intermediate scattering
function,
\ba
 S(\Delta {\bf K}, \omega) & = & \frac{e^{-\chi_l^2 \tilde{A}_1}}{\pi}
  \sum_{m,n=0}^\infty \frac{(-1)^{m+n}}{m!\ \! n!}
  \ \! \chi_l^{2(m+n)} \tilde{A}_3^m \tilde{A}_4^n e^{i(m-n)\delta}
 \nonumber \\
 & & \qquad \times
   \frac{\chi_l^2 \tilde{A}_2 + (m+n) \tilde{\eta}}
   {[\omega - (m-n)\tilde{\omega}]^2
    + [\chi^2 \tilde{A}_2 + (m+n) \tilde{\eta}]^2} .
 \label{dsfg}
\ea
This general expression clearly shows that both motions (diffusion
and T-mode) cannot be separated at all. The quasielastic or
Q-peak is formed by contributions where $m=n$. Analogously, the
contributions to the T-mode or T-peaks come from the sums with $n
\neq m$. If the Gaussian approximation is good enough, the value
of $\tilde{\eta}$ will not be too different from the nominal value
of $\eta$ and, therefore, both peaks will display broadening as
$\eta$ (or $\lambda$) increases. As an illustration, in Fig.
\ref{fig1} the widths of the Q-peak for the diffusion of Na atoms
on Cu(001) at two different coverages and two different surface
temperatures are plotted. As clearly seen, the experimental
\cite{andrew} and theoretical \cite{ruth6,ruth5} values of the
Q-peak width display broadening.
Taking into account the preceding discussion, this implies a faster
decay rate of the intermediate scattering function with the coverage
which should be attributed to a manifestation of the AZE.

\section{Conclusions}
\label{sec4}

The characteristic quantum short-time behavior of the observable
magnitude ($t^2$-law) is always present and, contrary to what is
widely accepted, this behavior is not always a clear-cut
characteristic of the QZE. In general, it depends on the type of
measurements carried out during the time evolution of the system.
This is true, at least, within this context (activated surface
diffusion of interacting adsorbates), where the evolution of the
intermediate scattering function in this time regime is governed
by friction-free (ballistic) motion. Moreover, within a Markovian
framework, the AZE in surface diffusion by corrugated surfaces has
been reported by experimental and theoretical works, but it has
not been recognized as such.

In this work, we have left on purpose for a future
investigation several important points in the surface diffusion
and how they are related to the QZE and AZE. First, the
diffusion mediated by tunneling. Second, the non-Markovian
character of the environment and therefore the role played by
the sub-Ohmic and supra-Ohmic regimes.
Third, at very short times, the adparticle is mainly inside any of the
potential wells of the corrugated
surface and the corresponding quantum motion is mainly intra-well
within a harmonic potential. This simple model allows us to
understand the physics associated with the so-called T-mode (or
frustrated translational mode). Recently, QZE and AZE
have been shown to occur for the undamped harmonic oscillator in
terms of the survival probability \cite{piilo}. A similar analysis
could be carried out in terms of the intermediate scattering
function.
Fourth, and finally, a true quantum mechanical calculation of
the surface diffusion by using the so-called stochastic
Schr\"odinger equation of the quantum theory of open systems
\cite{petruccione} could also be carried out. Work in these
directions is currently in progress.


\ack

Support from the Ministerio de Ciencia e Innovaci\'{o}n (Spain) under
Projects FIS2010-18132 and FIS2010-22082, and a joint Cuban-Spanish Project
(H.\ C.\ Pe\~nate-Rodr\'{\i}guez and  G.\ Rojas-Lorenzo) is acknowledged.
R.\ Mart\'{\i}nez-Casado thanks the Royal Society for a Newton Fellowship.
A.\ S.\ Sanz also thanks the Ministerio de Ciencia e Innovaci\'{o}n (Spain)
for a ``Ram\'on y Cajal'' Research Fellowship.


\end{document}